\documentclass{svmult}
\usepackage[utf8]{inputenc}
\usepackage{hyperref}
\usepackage{xcolor}
\usepackage{comment}
\usepackage[square,sort,comma,numbers]{natbib}
\usepackage{siunitx}
\usepackage{amsmath,amssymb}
\usepackage{geometry}
\usepackage{todonotes}
\usepackage{threeparttable}
\usepackage[misc]{ifsym}

\newcommand\aj{{AJ}}
\newcommand\apj{{ApJ}}
\newcommand\apjl{{ApJL}}     
\newcommand\apjs{{ApJS}}
\newcommand\aap{{A\&A}}
\newcommand\mnras{{MNRAS}}

\newcommand\pasa{{PASA}}

\DeclareSIUnit\parsec{pc}
\DeclareSIUnit\mag{mag}
\DeclareSIUnit\h{\textit{h}}
\DeclareSIUnit\days{d}

\newcommand{\beq}{\begin{equation}}
\newcommand{\eeq}{\end{equation}}
\newcommand{\bdm}{\begin{displaymath}}
\newcommand{\edm}{\end{displaymath}}

\begin{document}


\title{The Role of Type Ia Supernovae in Constraining the Hubble Constant}
\author{Dan Scolnic and Maria Vincenzi}
\institute{ Dan Scolnic (\Letter) \at Department of Physics, Duke University, Durham, NC, 27708, USA, \email{daniel.scolnic@duke.edu}
\and Maria Vincenzi (\Letter) \at Department of Physics, Duke University, Durham, NC, 27708, USA, \email{maria.vincenzi@duke.edu}}
%
%
\maketitle

\abstract{In the conventional / most studied local distance ladder measurements, Type Ia supernovae (SNe Ia) are used in two of the three rungs.    In the second rung, their luminosities are calibrated by standard candles like Cepheids or Tip of the Red Giant Branch (TRGB).  In the third rung, the high luminosities and standardizability allow SNe to be used to calibrate the `Hubble' relation between distances and redshifts.  Locally, the majority of distance ladder analyses find a high value of the Hubble Constant $H_0$ of $>70$ km/s/Mpc.  Given the discrepancy with the inferred value using CMB observations, great scrutiny must be given to the role supernovae play in measuring $H_0$.  Here, we review the main methodology, the many crosschecks for the supernova component of the distance ladder, and the various systematics studied.  We review the important role supernovae play to explain the small disagreements seen from various local analyses.  We also discuss analyses that employ an inverse distance ladder, which use similar sets of supernovae, but in the reverse direction, and yield a low value of $H_0$.  We conclude given all available evidence, it is difficult to find a way that a systematic in supernovae measurements, or a non-$\Lambda$CDM component of the universe which could be measured with supernovae, can help explain the Hubble tension. }


\section{Introduction}
When Edwin Hubble first measured the expansion of the universe, no supernovae (SNe) were used in this measurement \cite{Hubble29}. Instead, Hubble measured distances to galaxies by observing the Cepheids in the galaxy and Cepheids' properties discovered by Henrietta Leavitt \citep{Leavitt}. Today, both Cepheids and SNe are often used together to infer what is now called the Hubble constant $H_0$, which parameterizes the relation between expansion velocity and distance such that $v=H_0 d$. The Cepheids allow for a projection of a geometric scale, like that from parallax or megamasers, out onto the Megaparsec scale.  However, the range that Cepheids can be found and measured with the Hubble Space Telescope is limited to $\sim 40$ Mpc, where the measurements of motions of galaxies are still strongly affected by gravitational pulls of nearby galaxies (motions called peculiar velocities).  Therefore, SNe are needed as an extra rung along the distance ladder to reach out further into what is now deemed the `Hubble flow' ($\sim$100 to 600 Mpc), and precisely measure the expansion rate of the universe.

In the typical three rung distance ladder, SNe are used in the second and third rung.  In the second rung, their luminosities are calibrated with stellar distance indicators in the same galaxy.  In the third rung, their brightnesses calibrate the Hubble relation.  Ideally, one could remove the intermediate step, like Cepheids or Tip of the Red Giant Branch (TRGB), and go straight from geometric calibration to SNe.  Unfortunately, the rate of SNe in the local universe is not nearly frequent enough to provide multiple SNe in the few galaxies used for geometric anchors (the Milky Way, the Large Magellanic Cloud, NGC 4258).  Even with Cepheids or TRGB as a go-between, the low rate of SNe in the nearby universe (roughly one per galaxy per 100 years) is the limiting component of the precision of $H_0$ measurements \citep{Riess22}. At this point, the SH0ES team \citep{Riess22} for instance, utilizes every SN Ia that pass cosmological quality requirements and Cepheid suitability within $40$ Mpc.

 SNe Ia are used in the most cited local distance ladder analyses due to their high precision, large luminosity, and number of discovered sources.  However, as we will discuss in this review, studies have attempted to replace SNe Ia with hydrogen-rich core collapse SNe \citep{deJaeger22}, which while being less-precise standardizable candles, still provide an effective crosscheck.  Along the same lines, the role of SNe can be replaced with measurements of Surface Brightness Fluctuations \cite{Blakeslee21} or the Tully Fisher relation \cite{Schombert20}, a different type of standard candle that allows reach into the Hubble Flow.  With a similar purpose, there have been a tremendous number of re-analyses and studies of potential systematics of analyses of SNe Ia.  These SNe have been used in cosmological measurements for decades, so there is no shortage of checks on their accuracy.

 A related benefit of SNe Ia in particular is their usage for measurements of the expansion history, beyond just the current expansion rate.  There are multiple large surveys that have been charged over the last two decades to discover and measure SNe Ia in order to constrain properties of dark energy and the dark matter content of the Universe \cite{SNLS,Betoule14,Scolnic18,Jones19,Brout22}.  Leverage in understanding these properties comes from modelling the change of brightnesses of SNe Ia over a range of redshift (like within $0<z<2$).  As such, low redshift SNe are critically important, and the SNe that make up the third rung of the distance ladder are the same that provide the anchor for the relative distance constraints for measurements of expansion history \citep{Betoule14,Scolnic18,Brout22}.  Thereby, progress made in growing and understanding samples of low redshift SNe has been directly beneficial to both $H_0$ measurements of dark energy / dark matter measurements.

 For this review, we first discuss in Section~\ref{sec:2} the common three-rung path to constraining $H_0$ using SNe Ia.  We then discuss in Section~\ref{sec:3} the top systematic uncertainties in this measurement on the SN side, and how they have been quantified.  In Section~\ref{sec:4} and ~\ref{sec:5}, we present an accounting of the different crosschecks, using replacements or substitutions of the SNe sample.  In Section ~\ref{sec:6}, we discuss the implications of the inverse distance ladder approach. Finally, we present our discussions and conclusions in Section ~\ref{sec:7} and~\ref{sec:8}.

\section{The canonical path to $H_0$ with Type Ia Supernovae}
\label{sec:2}

We review here the formalism for deriving the Hubble constant with SNe Ia in the local distance ladder, as measured in \cite{Scolnic22} and \cite{Brout22} and used in \cite{Riess22}.  We follow the typical three-rung ladder, as shown on the left hand side of Figure 1.  As also shown in Figure 1, the third rung of the distance ladder is the low-redshift ($z<0.15$) part of the Hubble diagram, used to measure cosmological parameters like the equation-of-state of dark energy $w$.

We assume that a set of Cepheid or TRGB distances are calibrated with geometric measurements like parallax or megamasers.  This set of Cepheid or TRGB distances, expressed as $\mu_0$ here, can then be compared to SNe~Ia brightnesses $m_X$ in order to find a single offset $M_B$, which describes the absolute magnitude of a SN Ia. 

In most recent cosmological analyses with SNe Ia, the standardized brightness is measured with the Tripp formula such that 
\begin{equation}
m_X=m_B+\alpha x_1 - \beta c - \delta_{Bias} + \delta_{Host},
\end{equation}
where $m_B$, $x_1$ and $c$ are all independent properties of each light curve, $\alpha$ and $\beta$ are correlation coefficients that help standardize the brightness, $\delta_{Bias}$ is a correction due to selection effects and other biases as predicted by simulations, and $\delta_{Host}$ is a final correction due to residual correlations with host galaxy properties.  We call the standardized brightness $m_X$ instead of $m_B$ as in \cite{Riess22} to be clear that the brightness is standardized.  We note that typically SN Ia analyses like that in \cite{Brout22} will subtract off an absolute luminosity $M_B$ of SNe Ia to derive a distance modulus, but this must assume a fiducial $H_0$ value, which is what we are trying to derive.  

For a SN~Ia in the $i$-th Cepheid host, 
\begin{equation} 
m_{X,i}=\mu_{0,i}\!+\!M_B, \label{eq:snmagalt} 
\end{equation}
where $M_B$ is the fiducial SN~Ia luminosity, and $\mu_{0,i}$ is the distance derived from Cepheid measurements for each galaxy.

\ \par

The ladder is completed with a set of SNe~Ia that measure the expansion rate quantified as the intercept, $a_B$, of the distance (or magnitude)--redshift relation.  For an arbitrary expansion history and for $z>0$ as
\begin{equation} a_B=\log\,cz \left\{ 1 + {\frac{1}{2}}\left[1-q_0\right] {z} -{\frac{1}{6}}\left[1-q_0-3q_0^2+j_0 \right] z^2 + O(z^3) \right\} - 0.2m_X^0,\label{eq:aB} 
\end{equation}

\noindent measured from a set of SNe~Ia ($z, m_X^0$),  where $z$ is the redshift due to expansion, $q_0$ is the deceleration parameter, and $j_0$ is the jerk parameter. Typically, for $\Lambda$CDM, $j_0$ is set to 1.  The determination of $H_0$ follows from

\begin{equation} \log\,{\rm H}_0={0.2 M_B^0\!+\!a_B\!+\!5}. \label{eq:h0alt} 
\end{equation}

\ \par

The approach of \cite{Conley11} and \cite{Dhawan20} is now often used to account for covariance between rungs, as measurements of SNe Ia used in the second and third rung are correlated. Importantly, $q_0$ (and $j_0$) can only be constrained from SN Ia data, without the requirement of any additional information.

  \begin{figure*}
    \centering \vspace{20mm}
	\includegraphics[width=\textwidth]{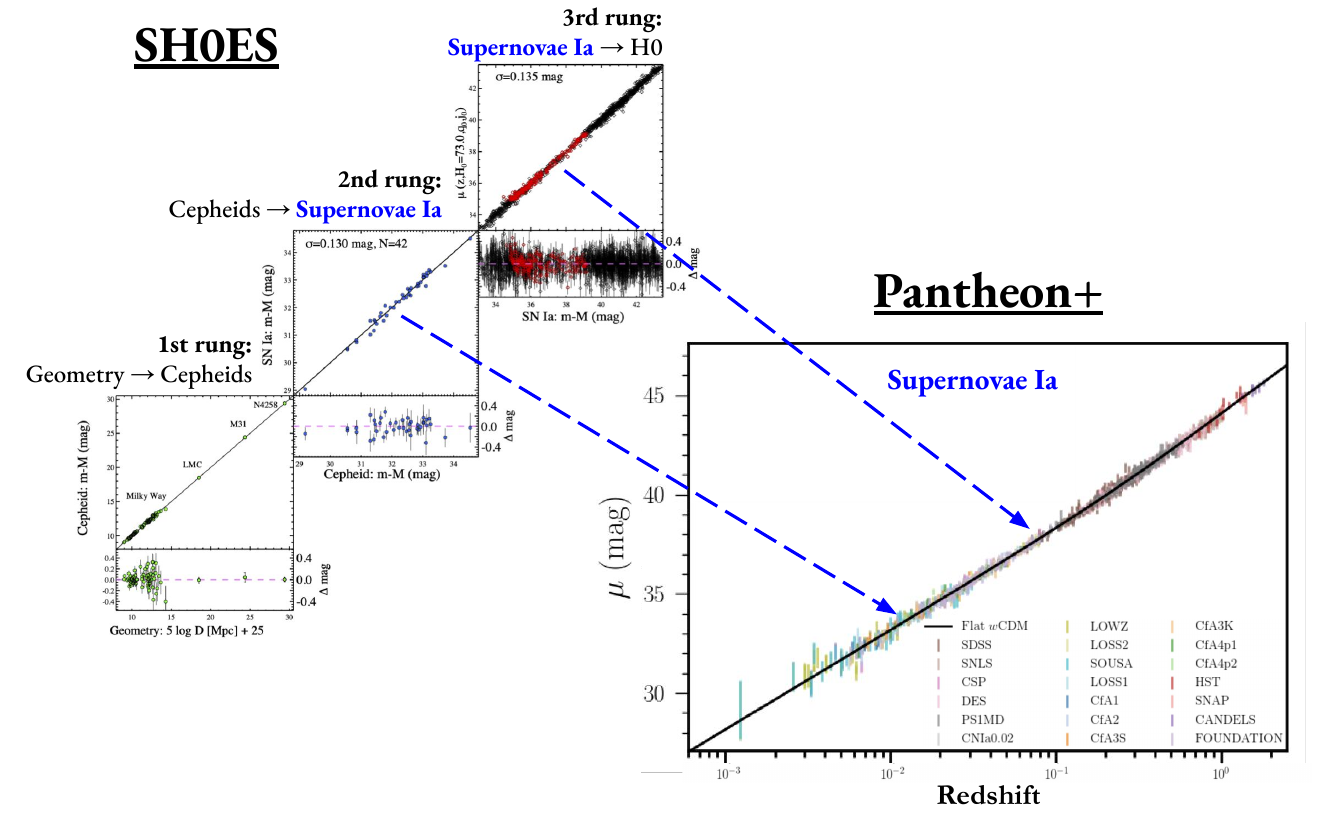} 
      \caption{\textbf{Left panel}: The `distance ladder' from \cite{Riess22}.   \textbf{Right panel}: The Pantheon+ ``Hubble diagram" from \cite{Brout22} showing the distance modulus $\mu$ versus redshift $z$. The SNe from the second rung make up the majority of the $z<0.01$ SNe in the Pantheon+ Hubble diagram, and the third rung makes up the rest of the $z<0.15$ SNe.  The uncertainties for $z<0.01$ shown on the Hubble diagram are much larger than those in the second rung, due to propagation of redshift uncertainties.}
    \label{fig:HD} \vspace{30mm}
\end{figure*}

\section{Top systematics on the path to $H_0$ with SN Ia}
\label{sec:3}

\begin{figure}
    \centering
    \includegraphics[width=0.95\textwidth]{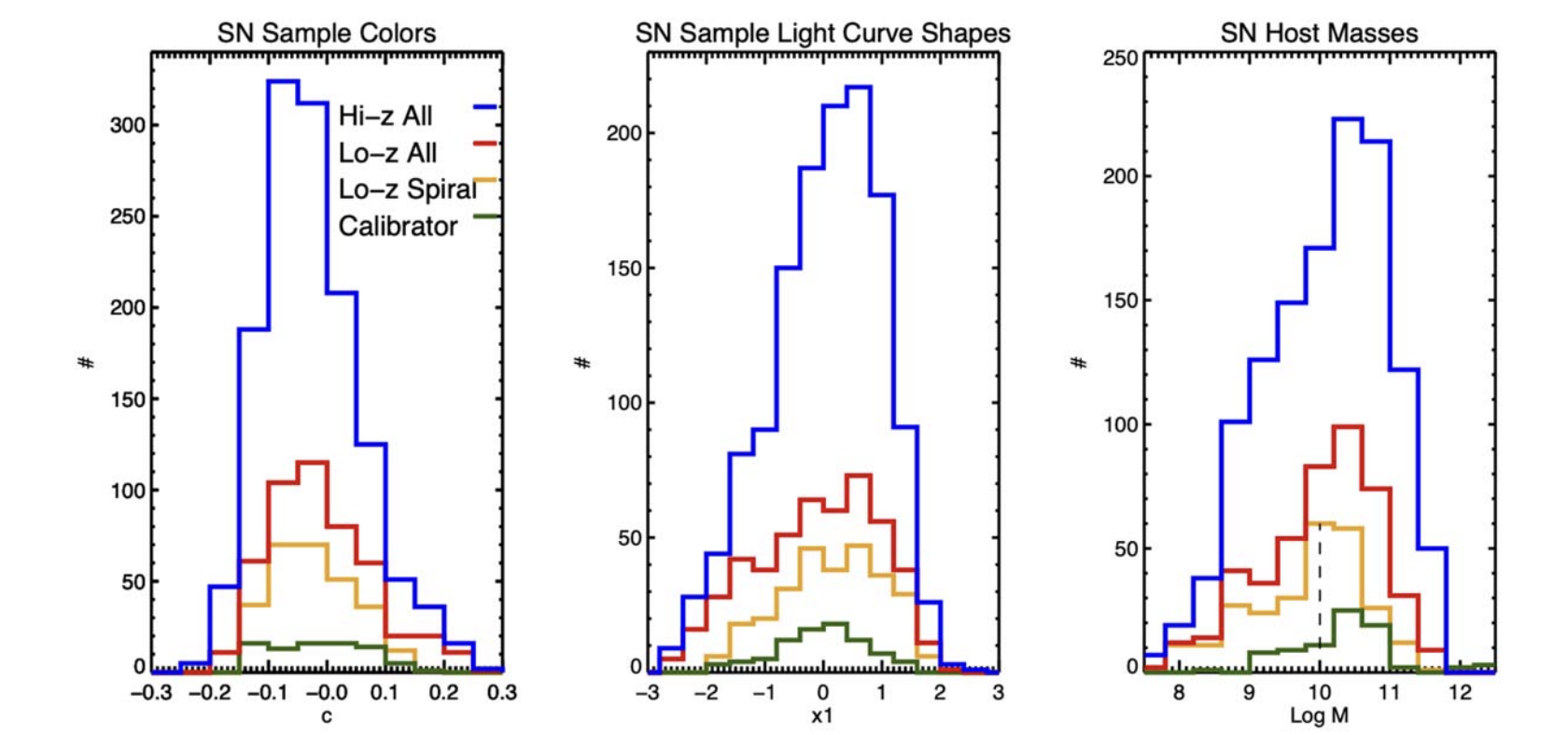}
    \caption{Adapted from \cite{Riess22}, a comparison of SN~Ia light-curve shape ($x_1$), colour ($c$) and the log host galaxy stellar mass (log $M_{\odot}$) for calibrator and Hubble flow samples (with different host galaxy type and redshift cuts).}
    \label{fig:r22_params}
\end{figure}

Cosmological measurements using SNe Ia are affected by various source of systematic uncertainties. However, measurements of $w$ using SNe are significantly more sensitive than measurements of $H_0$ with SNe. A simple explanation of the sensitivity is that $H_0$ is constrained by comparing SNe at $z\sim0.005$ to SNe in the Hubble flow at $z\sim0.05$, not too different in redshift and therefore similar in terms of SN properties and survey methods and telescopes used for SN observations (see Figure~\ref{fig:r22_params}). On the other hand, $w$ is constrained by comparing SNe at $z\sim0.05$ to $z\sim0.5$, in which evolution of SN properties is possible, and the surveys and telescopes used to find and measure the SNe will be quite different.  While new surveys like Foundation \cite{Foley18,Jones19}, ZTF \cite{Dhawan22} and DEBASS are attempting to measure more low-redshift SNe with the same telescopes as high-redshift SNe have been measured, it will take on the order of 10-20 years for these surveys to measure an equivalent amount of second-rung SNe ($\sim40$) as those in the current second rung. One route to overcome this hurdle is by extending the local volume for which host galaxy distances can be obtained, e.g. \citep{Jones2020}. 

\begin{figure*}
    \centering
\includegraphics[width=0.9\textwidth]{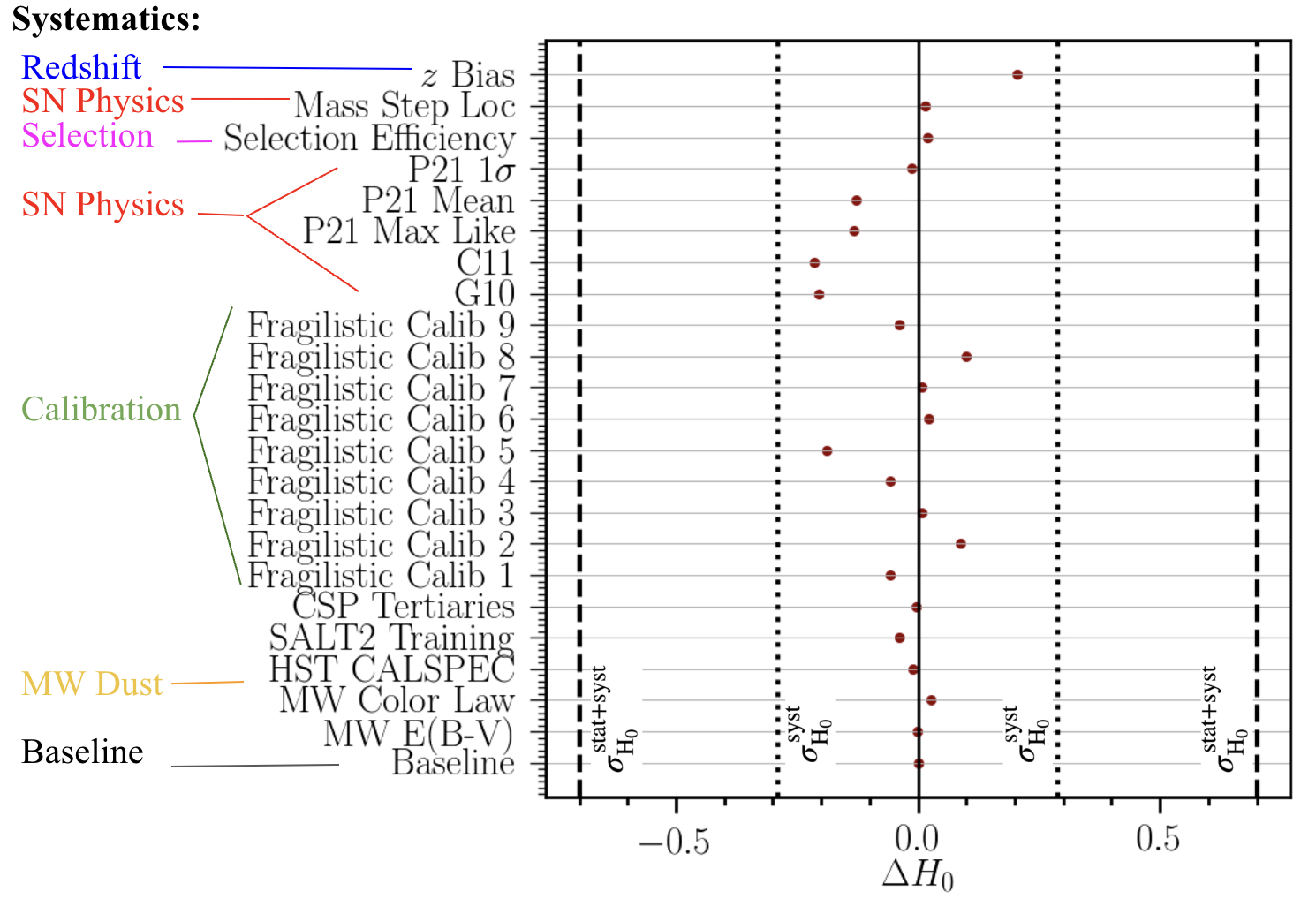}
\caption{Adapted from \cite{Brout22}, the impact on recovery of $H_0$ of the various systematic uncertainties tabulated. The units of these measurements are km/s/Mpc.  The dashed lines are given at $\Delta {\rm H}_0$ of 0.7, which is the entire contribution of the uncertainty in \cite{Riess22} from SN measurements. We add on labels explaining categories of systematic uncertainties.}
\label{fig:sysh0}
\vspace{.1in}
\end{figure*}

\begin{table}[]
 \centering
    \begin{tabular}{|p{3.7cm}|p{9.7cm}|p{2.3cm}|}
    \hline
     \textbf{Reference} &     \textbf{Notes}&                                          \textbf{Result (km/s/Mpc)} \\
    \hline
 \textbf{Specific systematic check for optical SNIa:} & ~ & ~ \\
    \cite{Murakami23} &  Uses spectral feature twinning process to improve standardization; checks dust modelling, intrinsic scatter modelling; & $73.01 \pm 0.92$ \\
    \cite{Peterson22} &  Checks different models of peculiar velocities / bulk flows & $\sigma_{H0}<0.2$ \\                                  
  \cite{brownsberger21} &  Checks SNIa Calibration by allowing individual SN survey offsets &                   $\sigma_{H0}<0.2$ \\  
         \cite{Jones18}  &    Checks impact of mass step, global vs local correlations &                        $\sigma_{H0}<0.15$  \\
       \cite{Burns18} &                  Checks light-curve fitting method; also does NIR fits &                                             $73 \pm 2$ \\
   \cite{Garnavich23} & Uses 4 rung distance ladder, checks SNIa host demographic systematic   &                         $74.6 \pm 0.9  \pm 2.7$  \\
      \cite{Dhawan22} &   Uses ZTF data alone, check on SNIa calibration   &                                        $76.94 \pm 6.4$  \\

   \textbf{  NIR SN Ia:} & ~ & ~ \\

   \cite{Dhawan18} &    Uses literature NIR SN (restframe $J$) and peak fitting   &                                         $72.8 \pm 2.8 $ \\
      \cite{Galbany22} &  Uses literature NIR SN (restframe $J$ and $H$ band)  &                       $72.3 \pm 1.4 \pm 1.4 $ \\

          \cite{Jones22} & Uses  RAISIN$\backslash$+literature NIR SN (restframe $Y$ band) and SNooPy fitting, check on dust & $75.9\pm2.2$   \\

      \cite{Dhawan23} &   Uses literature NIR SN and BayesN fitting &                     $74.82 \pm 0.97  \pm 0.84 $  \\
  
      \textbf{Removing SNIa:} & ~ & ~ \\

   \cite{Kenworthy22} &                        Eliminates SNIa rung entirely with 2 rung distance ladder &   $72.9^{+2.4}_{-2.2}$  \\
   
   \cite{deJaeger22} &   Uses SN II instead of SNIa &                                     $77.6^{+5.2}_{-4.8}$  \\
     
   \cite{Blakeslee21} &     Uses SBF standard candles instead of SNIa &                                  $ 73.3 \pm 0.7 \pm 2.4$  \\
    \cite{Kourkchi20} &                           Uses TF relation instead of SNIa  &                   $76.0 \pm 2.3\pm1.5$  \\
    \cite{Schombert20} &       Uses TF relation  &                   $75.1 \pm 1.1\pm2.3$  \\

\hline
\end{tabular}
    \caption{A summary of the various crosschecks and systematics on the supernova component of the distance ladder.  If two uncertainties are given, the first one is the statistical uncertainty and the second one is the systematic uncertainty.  }
    \label{tab:table1}
\end{table}

There have been a number of papers that have focused on systematics impact on $H_0$ and we list these in the top of Table~\ref{tab:table1}. \cite{Brout22} give a comprehensive overview of many of these systematics and how they may affect measurements of $w$ or $H_0$.  In Fig. \ref{fig:sysh0}, from \cite{Brout22}, we show the impact in units of $H_0$ for applying $1\sigma$ shift of each systematic. We group the systematics into various categories: Redshifts, SN Physics, Selection, Calibration, and Milky Way Dust.

Before discussing in more details each category of systematics below, it is important to understand which are the systematics $H_0$ is more sensitive to and, on the contrary, which are generally mitigated and have little impact on $H_0$.  
 The distance ladder is constructed such that the same probe is used in two of the three rungs (e.g. Cepheids in 1 and 2, SN Ia in 2 and 3). If a systematic is disproportionately affecting SN Ia in the second and third rung, or if it is introducing some substantial differences in the second rung and third rung SN Ia population, this will have a large impact on $H_0$. On the contrary, if a source of systematic is introducing some consistent offset that affects all SN Ia coherently, the impact of that systematic will be null due to the formalism described above. In other words, the offset will cancel and the effect of the systematic will be mitigated. 
 
\vspace{0.2cm}
 
\noindent \textbf{Calibration:} While calibration dominates the systematic error budget for $w$, its impact is small for $H_0$ because of the systematic mitigation discussed above. An illustration of this is given in \cite{brownsberger21} and \cite{Brout22c}. In particular, \cite{brownsberger21} check the possibility of `gray' calibration offsets per survey and finds a total uncertainty due to this effect no larger than 0.2 km/s/Mpc. They show that since many of the same surveys are used to measure SNe in the second and third rung of the distance ladder, the impact of potential biases in photometric calibration of the surveys will cancel. This is shown to not be the case when different surveys are used for the second and third rung.  This approach is done in \cite{Burns18} and \cite{Freedman19}, which while emphasizing the role of the CSP SNe, increases the sensitivity to calibration errors.

\vspace{0.2cm}

 \noindent \textbf{Redshifts:} $H_0$ measurements are more sensitive to redshift related systematic because redshift information is only used in one rung of the distance ladder (the third), so there is no systematic mitigation. \cite{Peterson22} explore a large number of models of bulk flows in the nearby universe, given various models that best improve the Hubble residual scatter from supernova measurements, and find that there could be changes in $H_0$ up to 0.2 km/s/Mpc. 
 They also find that the impact of including peculiar velocity corrections of the nominal method versus not including them is $\Delta H_0\sim0.5$. Peculiar velocities are discussed further in chapters of this book.  \cite{Carr22} check biases due in redshift measurements and find uncertainties on the level of 0.1 km/s/Mpc.   A predecessor study was done by \cite{Steinhardt22}, which similarly found small changes to $H_0$, though saw possible discrepancies in $\Omega_M H_0^2$ plane between subsamples of SNe for measured redshifts with different levels of precision.  No evidence for subsample differences was found by \cite{Carr22}.

\vspace{0.2cm}

\noindent \textbf{SN Physics and Selection:} 
SN intrinsic astrophysics and the role of SN dust remain one of the most poorly understood aspects of SN Ia cosmology. Recent analyses have investigated how systematics due to uncertainties in the physics of the supernova explosions and extragalactic dust extinction impact $H_0$ measurements \citep{Murakami23, P21_dust}. Most analyses have found that these effects have a small impact on $H_0$ because the differences between SN sub-populations selected in the second and third rung are not expected to be significantly different.

A good example of a potential systematic related to SN physics was discussed in \cite{Rigault15} following earlier SH0ES analyses (e.g. \cite{Riess11}).  \cite{Rigault15} showed evidence for a correlation between standardized brightness and the age of the host galaxy (quantified estimating the specific star-formation at the SN location).  In earlier measurements like \cite{Riess11}, the third rung of SNe had no galaxy-based selection applied, but the second rung, by tying to Cepheid discovery, favored star-forming galaxies. This would potentially lead to a bias in the recovery of $H_0$.  The size of the bias would depend on the relative differential fraction of host-galaxy demographics between the second and third rung multiplied by the size of the correlation.  Subsequent analyses \cite{Jones18} showed that this effect would likely be insufficient to explain the Hubble tension.  

Still, in the most recent SH0ES analysis \cite{Riess22}, the selection of SNe in the third rung of the distance ladder was done to be as similar as possible as the second rung. The SH0ES team only selected SNe found in star-forming galaxies which thereby removed the sensitivity to this systematic; the impact from this change was less than the statistical uncertainty from the supernova component of the distance ladder. Similarly, significant differences in dust extinction and/or color-related effects between SNe in the second and third rung could potentially bias $H_0$ measurements \citep{WojtakHjorth}. SN dust extinction and color-dependent corrections are encapsulated in the nuisance parameter $\beta$ (see Eq. 1). As a crosscheck, the $\beta$ parameter has been fitted separately in SN used in the second and third rung, and it has been found to be consistent \cite{Brout22}. This test, together with various NIR SN $H_0$ measurements \cite[][see Table~\ref{tab:table1}]{Dhawan18,Galbany22,Jones22,Dhawan23}, suggests that dust or color-dependent effects are not expected to significantly bias $H_0$, or to be a significantly underestimated systematic in current $H_0$ measurements.

In general, it is relatively simple to modify the SN sample selection in the second and third rung to ensure better consistency in terms of SN physics/dust and sample selection between rungs, and this approach has been implemented in the latest SH0ES analysis. 

\vspace{0.2cm}

\noindent \textbf{Other systematics:}
Finally, there have been various analyses \cite[e.g., ][]{Burns18, Murakami23} that implemented additional systematic tests, e.g., changing the light-curve fitter or adding spectroscopic information, but overall these appear to give very consistent (within $\sim0.3$ km/s/Mpc) results.  Similarly, isolating only to a single survey, like done in \cite{Dhawan23}, gives consistent results, though with much larger uncertainties due to the small number of calibrator SNe.

\section{Variants on the path to $H_0$ with Supernovae Ia}
\label{sec:4}
Some of the main crosschecks on the SNe Ia used for these analyses is varying the wavelength regime in which light curves are measured (i.e. optical to NIR) or the dataset used (i.e. the survey used to measure light curves).  As $H_0$ constraints are limited by the number of SNe found within 40 Mpc, there is considerable overlap in the data between these various studies.  The most popular path to check/improve the distance ladder with SN Ia is with measuring brightnesses from NIR light curves.  We list these papers in Table~\ref{tab:table1}.  Overall, even though the rest-frame band in which light curves are measured varies between these analyses, and the fitting method varies between these methods, there is generally very good agreement in recovered values of $H_0$.  One challenge multiple of these studies have found (e.g. \cite{Dhawan18}, \cite{Jones22}) is larger calibration offsets between samples than those found for optical studies.  A benefit of this type of study is the possibility of improved precision of distance measurements from NIR data, but the quality of older light curves has not typically been good enough to evaluate this possibility \cite{Peterson23}.  

\begin{figure*}
\includegraphics[width=\textwidth]{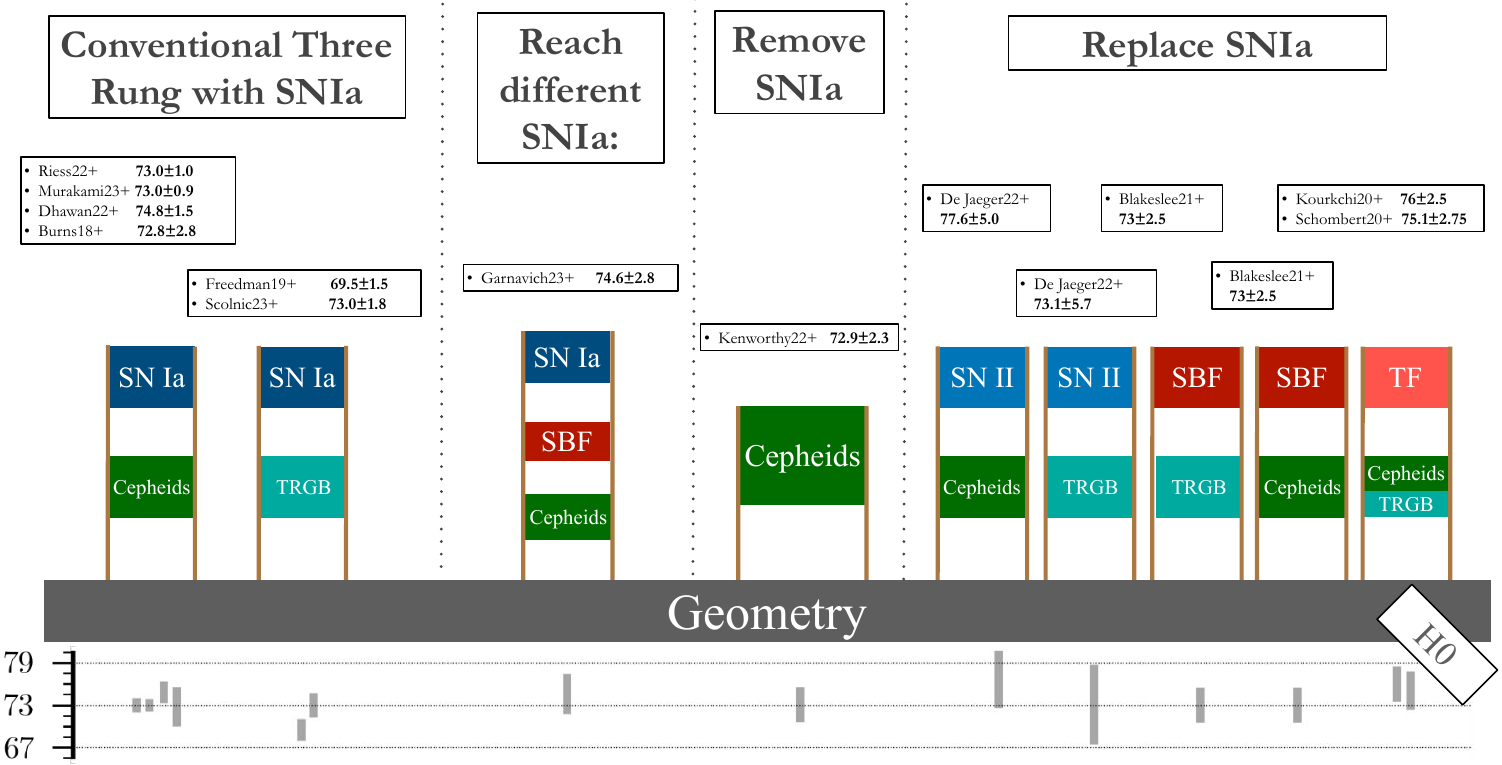}
\caption{A summary of papers and $H_0$ results given different uses and non-uses of SNe Ia. All $H_0$ measurements are in km/s/Mpc.}
\label{fig:garnavich}
\vspace{.1in}
\end{figure*}

An additional path is creating a ``4 rung distance ladder", as done in \cite{Garnavich23} and shown in Figure \ref{fig:garnavich}.  SNe Ia used in the SH0ES distance ladder are those found in late-type galaxies.  To avoid this specific subsample, one can add another rung in the distance ladder between TRGB/Cepheids and SNe - that from Surface Brightness Fluctuations (SBF).  The analysis of \cite{Garnavich23} improve on that of \cite{Khetan21}, which follow a similar method, but uses an inhomogeneous set of SBF measurements, which significantly increases the scatter of the tie between SBF and supernova measurements, and appears to bias $H_0$ to lower values. \cite{Garnavich23} find a value of $H_0=74.6\pm2.8$ km/s/Mpc, in good agreement with the SH0ES value.

\section{Removing Type Ia supernovae on the path to $H_0$}
\label{sec:5}
An alternative path is to remove SNe Ia altogether.  First, one can attempt to measure $H_0$ with a SH0ES-like distance ladder, but without the third rung; this is called a ``two-rung" distance ladder (see Figure~\ref{fig:garnavich}). This relies on knowing the redshifts to the nearby galaxies in which Cepheids are discovered. As these galaxies are mostly $z<0.01$, uncertainties in redshifts on the order of 300 km/s can contribute more than 10\% of the uncertainty per galaxy. Given the low signal-to-noise, any systematic biases from selection effects or peculiar velocity corrections may have a large effect on the measured value of $H_0$. \citep{Kenworthy22} attempt this with well-calibrated HST measurements of Cepheids from the SH0ES team, and measures $H_0=73.1\pm2.5$ km/s/Mpc.  A key component of this type of analysis is accounting for the inherent volumetric bias that there are more galaxies further away than nearby.

A separate path, instead of removing the third rung entirely, is to replace SN Ia with another type of standard candle. \cite{deJaeger22} attempt to replace SN Ia with SN II; while SN II are not typically considered standardizable candles, there has been good progress to improve their precision as distance indicators.  The scatter in the Hubble flow is $0.27$ mag, compared to the $\sim0.17$ mag found for SNe Ia.  The precision is enabled by leveraging a correlation between the luminosity and photospheric velocity and a color correction \cite{Hamuy02,deJaeger20}. The study finds weaker constraints on $H_0$, but good agreement with the SH0ES analysis: $H_0=77.6^{+5.2}_{-4.8}$\,km/s/Mpc.  

Another is using Surface Brightness Fluctuations, but in the same role as SNe Ia \cite{Blakeslee21}.  With those, they find $H_0=73.3\pm0.7\pm2.4$\,km/s/Mpc.  This combines first-rung calibration from Cepheids and TRGB, though they yield similar results ($H_0=73.44$ for Cepheids and $H_0=73.20$\,km/s/Mpc for TRGB, each with $\sim5\%$ total uncertainty). The number of calibrators is just 7, compared to the 42 that SH0ES uses, thereby limiting the overall precision.  Additionally, the Tully-Fisher (TF) relation can be done in this third-rung role.  \cite{Schombert20} and \cite{Kourkchi20} calibrate the TF relation with Cepheids and TRGB; after doing so, \cite{Schombert20} recovers $75.1\pm 2.3 (stat.) \pm1.5 (syst.)$\,km/s/Mpc and \cite{Kourkchi20} finds $76.0 \pm 1.1(stat.) \pm 2.3(syst.)$\,km/s/Mpc.

\section{Inverse Distance Ladder to $H_0$ with Supernova}
\label{sec:6}
The same set of SNe Ia that are calibrated to a physical distance scale in the SH0ES distance ladder, can also be used as un-calibrated relative distance indicators to constrain $H_0$ in combination with other probes. Here we describe the process of combining the un-calibrated SNe Ia with constraints from Baryon Acoustic Oscillation (BAO). BAO constraints on the expansion history $H(z)$ and that extrapolate $H_0$ have been shown to achieve low values of $H_0$ \cite{Macaulay2019,Feeney19}. We note that such constraints 1) assume the sound horizon from CMB constraints and 2) are model dependent because they assume $\Lambda$CDM to infer $H(z=0)$ from BAO datasets at $z\sim0.5$ and do not allow for late-time physical solutions. However, the SNe Ia can be used to solve the latter of these two. 

Instead of calibrating the SN Ia intrinsic magnitude to the distance ladder, they can be calibrated at the typical BAO redshift to the distance scale set by BAO and for which there is substantial overlap in $z$ of many SNe Ia in current datasets. Having calibrated the SNe Ia to the BAO scale at moderate redshift, one can use the SNe Ia to constrain the expansion history without assuming $\Lambda$CDM to infer $H_0$ (see smoothed blue curve of Figure \ref{fig:HD}). In this case, one finds a very similar result ($H_0=68.57\pm0.9$) to that of the CMB under the assumption of $\Lambda$CDM (grey curve of Figure \ref{fig:HD}). We note here that using SNe Ia calibrated to the BAO, which itself obtains the physical scale from the sound speed in the early universe, to recover a low value of $H_0$ has spurred many discussion to understand the impact of the value of the sound horizon.

  \begin{figure*}
    \centering \vspace{20mm}
	\includegraphics[width=.72\textwidth]{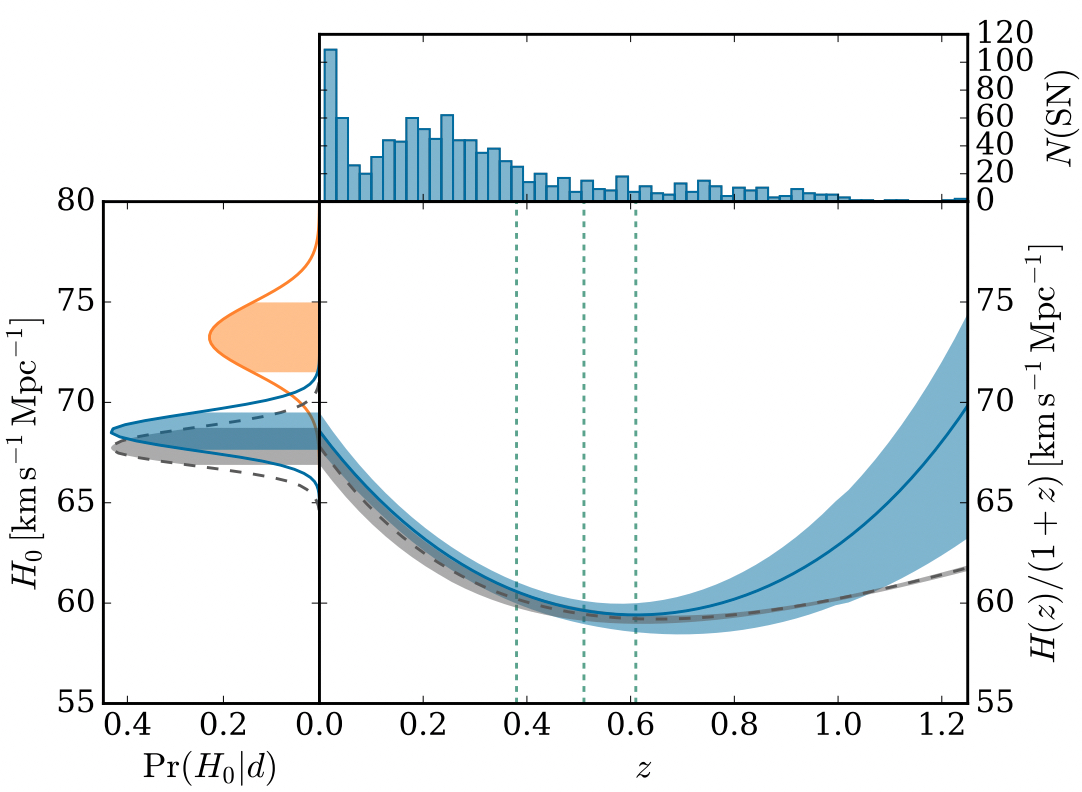} 
      \caption{From \cite{Feeney19}, history for BOSS BAO, Pantheon SNe and \textit{Planck} $r_d$ assuming smooth expansion
and early-time physics only (blue), and for Planck assuming $\Lambda$CDM (grey). BAO redshifts are shown as short-dashed
lines. Left panel: corresponding $H_0$ posteriors and Cepheid
distance ladder measurement (orange). Top panel: redshift
distribution of Pantheon SNe}
    \label{fig:HD} \vspace{30mm}
\end{figure*}

\section{Discussion} \label{sec:Discussion}
\noindent\textbf{Improving measurements of $H_0$ in the future:}\\
\label{sec:7}
While constraints on $w$ should easily improve with upcoming large SN samples, the road to improving constraints on $H_0$ is more challenging. {There are a limited number of SNe~Ia that will explode in the near future within a $\sim40$\,Mpc radius, a constraint due to {\it HST} discovery limits of Cepheids.} At roughly one to three SN~Ia per year, it will likely take more than one decade to double the current sample of $42$ SNe calibrated by SH0ES Cepheid hosts. An exciting possibility is that the distance range for precise measurements of Cepheids or TRGB can be extended.  As the volume of discovered SNe roughly increases with distance cubed, then even a $25\%$ increase in the range of Cepheids or TRGB measurements would allow a doubling in the number of usable SN Ia in the second rung of the distance ladder.  This may be a possibility with new telescopes such as the James Webb Space Telescope, or further into the future, the Nancy Grace Roman Space Telescope, or ground-based Extremely Large Telescope (ELT).  

The other path towards improving the constraint from SN Ia is to improve the precision of the measurements.  This is the path followed by papers like \cite{Murakami23} which tried using spectral features to improve the standardization, or the large number of papers that measure SNe Ia in the NIR.  \cite{Murakami23} are able to improve the scatter of SN Ia from 0.14 to 0.12 mag, which is similar to increasing the number of SN Ia by $\sim35\%$.  The main challenge of this approach is that the benefit of the new idea depends on how much data is available for its application in past literature measurements.  New types of measurements can be made for future nearby supernovae, but we can not re-measure past SN Ia.\\

\noindent \textbf{The role of SNe Ia when comparing local probes:}\\
Within the larger discussion of tension between early and late universe probes, there has been various discussions of smaller tensions between late-universe probes. One of note has been potential discrepancies when using Cepheids or TRGB in the distance ladder.  While a lot of this focus has been on Cepheids versus TRGB, we note that often the SNe that are used to project out the distance ladder can cause discrepancies.  An interesting example of that is shown from \cite{Scolnic23}, which breaks down changes in $H_0$ from the CATS and CCHP teams \cite{Freedman19}.  As shown in Table 1.2, the CATS team explains that SN choices account for more than 60\% of the difference with other analyses, whereas the TRGB side is less than 40\%.  A number of these choices have been discussed in this review, including the application of peculiar velocity corrections and consistency amongst surveys used for the different rungs of the distance ladder.

\begin{table*}
    \small
    \centering
    \caption{Sources of Differences in $H_0$ Between TRGB analysis by CATs \citep{Scolnic23} and CCHP \citep{Freedman19} (in units of $H_0$)}
\label{tb:trgbdiff}
    \begin{tabular}{lc}
        \hline
        \textbf{Term} & $\Delta$CCHP \\
        & (km/s/Mpc)  \\
        \hline
        \multicolumn{2}{c}{\textbf{SN Related}} \\
        \hline
        1. Include SN 2021pit, 2021rhu, 2007on  & 0.6 \\
        2. No TRGB detected in N5584, N3021, N1309, N3370 & 0.0 \\
        3. Peculiar Flows (Pantheon$+$) & 0.4 \\
        4. Hubble Flow Surveys (Pantheon$+$) & 1.1 \\
        \hline
        \textbf{SN subtotal} & \textbf{2.0} \\
        \hline
        \multicolumn{2}{c}{\textbf{TRGB Related}} \\
        \hline
        5. Fiducial TRGB Calibration/Tip-Contrast Relation & $1.4$ \\
        \hline
        \textbf{Total} & \textbf{3.4} \\
        \hline
    \end{tabular}
    \begin{tablenotes}
            \item\textbf{Comments:} Adapted from \cite{Scolnic23}, this shows how differences that appear to be from anchor measurements of Cepheids versus TRGB are more due to differences in the supernova analysis. Here, $\Delta$CCHP= differences between \cite{Freedman19}.  Descriptions of individual entries: (1) CCHP did not include the two SNe from 2021 and excluded SN 2007on, (2) CATs did not detect the TRGB in these four most-distant SN host galaxies; (3) Pantheon$+$ accounts for peculiar motions which produce a highly significant improvement in the dispersion of the Hubble diagram (see \citealp{Peterson22})  (4) CCHP measures the Hubble flow from a single SN survey which has an offset with respect to the mean of many surveys in Pantheon$+$ as shown in \cite{brownsberger21}; (5) This term is difference in the calibration of TRGB from NGC 4258 as applied to SN hosts and is discussed by \cite{Scolnic23} (see Section 4).
    \end{tablenotes}
\end{table*}

\section{Conclusion}
\label{sec:8}

We discuss the various systematic uncertainties of as well as the crosschecks on the SN Ia component of the distance ladder, which occupies two of the three rungs.  We find generally that systematics in the measurement of $H_0$ from the SNe are at the scale of 0.3\,km/s/Mpc. The range of $H_0$ values recovered from crosschecks is from $72<H_0<78$\,km/s/Mpc.  Therefore, it appears unlikely that SN Ia are artificially causing a high value of $H_0$ to be found with the local distance ladder.  Still, additional studies of SN Ia are important in order to compare other parts of the distance ladder, like TRGB versus Cepheids, as the sample overlap does not always yield high significance from a direct comparison.

\begin{acknowledgement}
 D.S. thanks the John Templeton Foundation for their support of grant \#62314. D.S. is supported by DOE grant DE-SC0010007, the David and Lucile Packard Foundation.

M.V. is supported by NASA through the NASA Hubble Fellowship grant HST-HF2-51546.001-A awarded by the Space Telescope Science Institute, which is operated by the Association of Universities for Research in Astronomy, Incorporated, under NASA contract NAS5-26555.
\end{acknowledgement}

\end{document}